\documentclass[aps,prl,showkeys,showpacs,preprint,groupedaddress]{revtex4}
\usepackage{graphicx}

\begin{document}

\title{Quantum correlations, measurement and the origin of
uncertainty}

\author{Daniele Tommasini}

\affiliation{Departamento de F\'\i sica Aplicada, \'Area de
F\'\i sica Te\'orica, Universidad de Vigo, 32004 Ourense, Spain}

\email[]{daniele@uvigo.es}

\date{\today}

\begin{abstract}
I argue that the correlations that are predicted by Quantum Field
Theory should not be interpreted as a real sign of non locality.
\end{abstract}

\pacs{11.10.-z; 03.65.Ta}

\keywords{Quantum Field Theory; Measurement Problem;
Einstein-Podolsky-Rosen paradox; Entanglement}

\maketitle

In 1935 Einstein, Podolsky and Rosen (EPR) \cite{EPR} pointed out
that Quantum Mechanics apparently implied some mysterious,
instantaneous action at a distance. This was considered a paradox,
since it made the Quantum Theory incompatible with Special
Relativity, suggesting the need for a more fundamental description
of Nature.

Recently, I have shown that this EPR paradox is removed in Quantum
Field Theory (QFT) due to an uncertainty on the number of soft
photons that can appear in any process \cite{QEDEPR,EPRsolve}. For
instance, let me consider a typical EPR experiment, involving two
spin 1/2 particles A and B that are produced in coincidence in a
rotation-invariant process with zero initial angular momentum. In
my previous works, I have shown that when the spin component is
measured on particle A, no certain conclusion can be obtained for
the result of the measurement on B, since the angular momentum
conservation involves also an unknown number of soft photons
\cite{QEDEPR,EPRsolve}. This fact is sufficient to remove the
paradox at least in practice, since it makes it impossible to {\it
use} the supposed quantum nonlocality, e.g.˜ to teleport a state
to distant particles.

In Ref.˜ \cite{QEDEPR}, I have also argued that the QFT
correlations turn out to be smaller than those obtained by
ignoring the soft photons, but still can allow for the observed
violation of the ``Bell's inequalities" \cite{Bell,BJ}. However,
such correlations are commonly thought to be a sign of quantum
nonlocality by themselves. Then, even if the hardest part of the
EPR paradox is removed by the soft photons argument, we are left
with the residual mystery of explaining how comes that in the
correlations the angular momentum is exactly conserved after the
measurement due to an apparent cancellation amongst distant
particles (A, B and the soft photons).

In Ref.˜ \cite{EPRsolve}, I have noticed that in QFT the angular
momentum is conserved locally. Therefore, if one interprets the
measurement as the result of scattering processes, only the
particles that come in causal contact with A during the
measurement can change their state as a consequence of the
measurement. No effect can be induced on the distant particles (B
and the possible soft photons), and the correlations should be
causal. This is also a sound prediction of QFT \cite{Weinbooks},
however it seems to be contradicted by the fact that in the actual
correlations \cite{QEDEPR} the angular momentum is conserved for
the system of particles A, B and the soft photons, independently
of the measuring apparatus. How comes that some quantities that
are obtained by a causal theory apparently imply an instantaneous
action at a distance (whatever unobservable it would be)?

Even if this can be considered a philosophical problem, I think
that it is relevant. One could hope that it would disappear in a
more fundamental theory, such as a String Theory. Of course, such
an attitude is legitimate, and is justified by the fact that QFTs
themselves are not thought to be the ultimate Theory of
Everything, since e.g.˜ they do not describe gravity. However, I
think that we should already get a solution to this apparent
contradiction without leaving QFT. In fact, it has been shown that
QFT is the form that any Relativistic Quantum Theory should get at
``low" enough energies \cite{Weinbooks}. Since our problem has to
do with ``large" distances, i.e.˜ with low energies, it can be
expected that it should be solved without waiting for the ultimate
Theory of Everything. In this brief note, I will show that this is
actually the case: when QFT is interpreted correctly, the
contradiction between the causality of the theory and the apparent
nonlocality of the correlations that it predicts will disappear.

To see this, let me first notice that the field equations are
deterministic. If we consider the Universe as a whole, instead of
dividing it in an examined system, a measurement apparatus and the
rest, QFT would just describe a smooth, deterministic evolution.
We can consider as the dynamical variables the fields themselves,
or equivalently the Complete Set of Commuting Observables (CSCO)
made with the number operators $N^i(\vec p,\lambda)$ that count
the (density) number of particles of the type (i) (e.g.,
electrons, or positrons, or photons, etc.) having momentum $\vec
p$ and spin/helicity $\lambda$. In any case, in the Heisemberg
picture, all these dynamical variables have a well determined,
smooth evolution (equivalently, in the Schr\"odinger picture the
vector state evolves continuously in a deterministic way). Then,
where comes the quantum uncertainty from? It is obvious that the
determinism of the theory is not destroyed by the apparatus of
measurement itself, that is included in the system of the whole
Universe. The uncertainty only appears when we split the world in
a measuring apparatus and a measured system (and the rest). Of
course, this division is necessary if we want to do physics, i.e.˜
to describe the phenomena that occur in a given subsystem of the
Universe.  But it is important to understand that this is the
unique origin of the uncertainty: the separation between the
object and the observer, and the limited point of view of the
latter, that has to renounce even to take into account the effect
of the measurement on the measuring apparatus itself.

This artificial division implies a lack of information which is
the ultimate reason for the probabilistic character of the
predictions of the QFT. This is also evident in scattering theory:
the scattering matrix (giving the correlations) is obtained by
considering the amplitudes of the state at $t\to\infty$ (using the
Schr\"odinger picture) into the out states, that are just the
possible eigenstates of the CSCO $\{N^i(\vec p,\lambda)\}$. Any
such a scalar product is usually thought to correspond to a
``collapse" process, but the latter is actually a result of smooth
scattering processes. In other words, the state vector of the
Universe evolves continuously, and the correlations are just made
out of its components, integrating out the variables that do not
correspond to the observed (sub)system \footnote{Actually, there
is also a theoretical impossibility to {\it define} unambiguously
the system, due e.g.˜ to the uncertainty on the soft photons.}.
Therefore, the collapse does not happen at a fundamental level.
According to the discussion of Ref.˜ \cite{EPRsolve}, this is
sufficient to remove also the last piece of the EPR paradox, i.e.˜
the contradiction between the causality of QFT and its apparently
nonlocal correlation functions. It is now obvious that the
correlations are actually the reminder of the common origin of our
two particles A and B, as we also expected from the causality of
the theory. This also reconciles us with the relativistic
intuition.

In my previous paper \cite{EPRsolve}, I concluded that Einstein
was right in thinking that local realism was a fundamental
characteristic of any reasonable theory of physics. Moreover, we
see now that, in some sense, his intuition that the solution to
the EPR paradox should be a deterministic theory was also correct.
QFT is amazing, since it is at the same time extremely
probabilistic and fully deterministic, without being a Hidden
Variable theory \footnote{Actually, it could be said that the
number of soft photons plays a role similar to that of an Hidden
Variable.}. The Universe evolves in a deterministic way, while the
act of measuring on an object subsystem, using another subsystem
called measuring apparatus, implies a restricted knowledge,
allowing only for probabilistic predictions. Einstein would
probably say that God does not play dice; but we actually do.

I am grateful to Humberto Michinel, Ruth Garc\'\i a Fern\'andez,
Uwe Trittmann and Rafael A. Porto for very useful discussions,
help, comments and criticism.

\bibliography{Correl}

\end{document}